\documentclass{article}

\usepackage{PRIMEarxiv}
\usepackage{longtable}
\usepackage[utf8]{inputenc} 
\usepackage[T1]{fontenc}    
\usepackage{hyperref}       
\usepackage{url}            
\usepackage{booktabs}       
\usepackage{amsfonts}       
\usepackage{nicefrac}       
\usepackage{microtype}      
\usepackage{lipsum}
\usepackage{fancyhdr}       
\usepackage{graphicx}       
\graphicspath{{media/}}     
\usepackage{txfonts}
\usepackage{xcolor}

\title{Estimation of the solar wind extreme events
\thanks{\textit{\underline{Citation}}: 
\textbf{Larrodera, C., L. Nikitina, and C. Cid (2021). “Estimation of the solar wind extreme events”.
In: Space Weather. DOI:https://doi.org/10.1029/2021SW002902.}} 
}

\author{
  C. Larrodera \\
  University of Alcalá \\
  Alcalá de Henares\\
  \texttt{carlos.larrodera@edu.uah.es} \\
  \And
 L. Nikitina \\
  Canadian Hazards Information Service \\
  Ottawa\\
  \And
  C. Cid \\
  University of Alcalá \\
  Alcalá de Henares\\
  \texttt{consuelo.cid@uah.es} \\
}

\begin{document}
\maketitle

\begin{abstract}
This research provides an analysis of extreme events in the solar wind and in the magnetosphere due to disturbances of the solar wind.
Extreme value theory has been applied to a 20 year data set from the Advanced Composition Explorer (ACE) spacecraft for the period 1998-2017. The solar proton speed, solar proton temperature, solar proton density and magnetic field have been analyzed to characterize extreme events in the solar wind. The solar wind electric field, vB$_{z}$ has been analyzed to characterize the impact from extreme disturbances in the solar wind to the magnetosphere. These extreme values were estimated for one-in-40 and one-in-80 years events, which represent two and four times the range of the original data set. The estimated values were verified by comparison with measured values of extreme events recorded in previous years. Finally, our research also suggests the presence of an upper boundary in the magnitudes under study.
\end{abstract}


\section{Introduction}

Scientific progression in the last decades has made modern society more dependent on technology. Due to the interdependence between different components of technological infrastructure, a severe space weather event could cause a cascading effect in different aspects of modern life, from disruption in electric power grids to spacecrafts malfunction and navigation problems. For example, one of the largest magnetic storms of the last century, occurring in March 1989, caused widespread effects in the Hydro-Québec power system in Canada (see, e.g., \cite{Boteler_2019_Space_weather}).
\cite{Riley_2018_Extreme_Space_Events} state that the cost of a worst-case scenario 1-in-100 years magnetic storm would include: (1) 1–2 Trillion USD dollars of economic loss; and (2) 130 million people without electrical power for several years, based on the destruction of several hundred transformers. 

The main driver of geomagnetic storms is the solar wind, hence, knowledge of the most severe disturbances in the solar wind is essential to both forecast and potentially mitigate risks related to space weather events. 
Extreme value theory (EVT) is a statistical method developed to analyze the likelihood of occurrence of rare and severe events (see \cite{Gumbel_1958_extreme, Coles_2001_Extreme_Book} and references therein). This theory has been applied in different fields, from hydrology and meteorology (see, e.g. \cite{Gumbel_1958_extreme}), to finance \cite{Embrechts_1994_Extreme_Events_Finance}, and public health \cite{Thomas_2016_extreme_public_health}. 
In recent decades, extreme value theory has been applied to estimate extreme values in different aspects of solar physics and space weather. In particular, extreme value analysis has been applied to the study of extreme geomagnetic storms (\cite{Siscoe_1976_extreme_events, Chen_2019_Extreme_Events,Elvidge_2020_Extreme_value,Love_2015_extreme_dst,Love_2019_Extreme_values, Nikitina_2016_Extremes,Thomson_2011_Extreme_values}), solar energetic proton flux (\cite{Koons_2001_extreme_values,Ruzmiakin_2011_extreme_values}), the electron flux in the outer belt of the magnetosphere (\cite{OBrien_2007_extreme_events}), and to the analysis of the solar cycle \cite{Asensio_2018_Extreme_value, Acero_2018_Extreme_sunspots}.  

In the present study extreme value theory is applied to estimate extreme values of solar wind characteristics like the interplanetary magnetic field magnitude, the solar proton speed, the solar proton temperature and solar proton density along with other magnitudes like the solar wind electric field, vB$_{z}$, to characterize the response of the magnetosphere to the extreme events in the solar wind. EVT also provides an opportunity to estimate the return value of these magnitudes, expected one-in-40 and one-in-80 years.

The paper is structured as follows: Section \ref{sec:extreme_values} details the statistical approach used in the analysis.
Section \ref{sec:data} describes the data set and it's temporal resolution.
Section \ref{sec:solar_wind} shows EVT applied to different magnitudes of the solar wind. Finally, Section \ref{sec:conclusions} shows the results and conclusions of the research.
 
\section{Extreme value theory} \label{sec:extreme_values}

Extreme value theory (EVT) is a statistical method developed to analyze the likelihood of the occurrence of severe events, i.e. events with a low probability of occurrence. For this analysis the whole data set has been separated into blocks of the same duration, and then the EVT has been used to analyze the statistical behavior of the maximum values $M_{n}$ for each block of data $X_{1}\dots X_{n}$, corresponding to a certain period of time \cite{Coles_2001_Extreme_Book}.
\begin{equation}
    M_{n}=\max \left(X_{1},X_{2},\dots X_{n} \right),
\end{equation}
where $M_{n}$ are assumed to be independent and identically distributed. 

The distribution of the maximum values ($M_{n}$) can be described by the Generalized Extreme Value (GEV) distribution which is defined by the cumulative probability ($p$) and depend on three parameters: the location parameter ($\mu$), the scale parameter ($\sigma$), understood as the variance of the data, and the shape parameter ($\xi$) which defines the behavior of the tail of the GEV distribution. Depending on the sign of the shape parameter, the GEV distribution has three different forms:

\renewcommand{\labelitemi}{\textbullet}
\begin{itemize}
\item $\xi>0 \Rightarrow$ Fréchet distribution,
\item $\xi=0 \Rightarrow$ Gumbel distribution,
\item $\xi<0 \Rightarrow$ Weibull distribution,
\end{itemize}
where the probability function for this distribution is: 
\begin{equation}
    p\left(M_n <z \right)=\exp\left(-\left[1+\xi \left(\frac{z-\mu}{\sigma}\right)\right]^{-1/\xi}\right),\mbox{ if \space} \xi \neq 0,
    \label{eq:GEV}
\end{equation}
\begin{equation}
    p\left(M_n < z\right)=\exp\left(-\exp \left(\frac{z-\mu}{\sigma}\right)\right), \mbox{ if } \xi = 0.
    \label{eq:GEV2}
\end{equation}

Large disturbances in the solar wind produce severe space weather events that can last from hours to several days \cite{Gopalswamy_2016_CMEs}. Therefore, it is important to define the size of the blocks in order to capture the severe events and assure that they are independent.
A block size of two days has been chosen for the extreme value analysis, checking that the $M_n$ values are related to different space weather events. Extension of the block size to larger amount of days should provide approximately the same fitting results, but 2 days were chosen for the analysis to keep larger data set and get more precise confidence intervals for extreme value estimation. Indeed, \cite{Zhang_2008_ICME_size} estimate an average size of 41.2 hours for the ICMEs. In the case that two consecutive points have been selected, the lower one was discarded, assuring only one point per event.

In order to choose the proper distribution function (Fréchet, Gumbel or Weibull) and perform a regression analysis, special coordinates which transfer the GEV distribution to a straight line were used.
The coordinates for the regression analysis are the double logarithm of the probability $-\ln\left(-\ln(p)\right)$ versus $z$ for the case $\xi = 0$ (Eq. \ref{eq:GEV2}) and the double logarithm $-\ln\left(-\ln(p)\right)$ versus $log(z)$ for the case $\xi \neq 0$ (Eq. \ref{eq:GEV}) \cite{Coles_2001_Extreme_Book}. The analyzed data set was fitted to both distributions (Eq.\ref{eq:GEV} and Eq.\ref{eq:GEV2}) to obtain the shape parameter ($\xi$) from the fitting and the R-Squared ($R^{2}$) to decide the appropriate distribution and proceed to further analysis.

The estimated return value ($z$) for a specific return period ($T$) can be calculated from the EVT using the probability $p(M_n<z)$:

\begin{equation}
    T=\frac{1}{1-p\left(M_n<z\right)}.
    \label{eq:return_period}
\end{equation}

\section{Data sample and temporal resolution} \label{sec:data}

In this analysis, level 2 data from 1998 to 2017 from the Advanced Composition Explorer (ACE) spacecraft and is located at the Lagrange point L1 were used. On one hand, the data set contains the solar proton speed (v) (with an energy range between 0.5 and 100 KeV), the solar proton density (N) and the solar proton temperature (T) from the Solar Wind Electron Proton Alpha Monitor 
(SWEPAM/ACE) \cite{McComas_1998_ACE_SWEPAM} and the Solar Wind Ion Composition Spectrometer (SWICS/ACE) \cite{Gloeckler_1998_SWICS} with a temporal resolution of 12 minutes. On the other hand, the interplanetary magnetic field magnitude (B) and Z component (B$_{z}$) from the magnetometer (MAG) \cite{Smith_1998_ACE} with a temporal resolution of 4 minutes were chosen.
The MAG and SWEPAM instruments provide higher temporal resolutions, 1 and 16 seconds for MAG and 64 seconds for SWEPAM, but we decided to use 4 and 12 minutes temporal resolution respectively for this task.
The data coverage is near 100\% during the whole period of 20 years of analysis, which coincides with the first 20 years of life of the ACE mission, launched in August 1997. This 20-year period allows one to establish return periods that are two and four times the duration of the data set, i.e., one-in-40 and one-in-80 years.
In order to study the response of the magnetosphere to the disturbances created by the solar wind, the solar wind electric field ($E=v|B_{z}|$) has been analyzed. 
The largest values considered in our study has been set by the percentile 99 of the cumulative distribution function of each magnitude.

\begin{figure}[!htb]
	\includegraphics[width=\columnwidth]{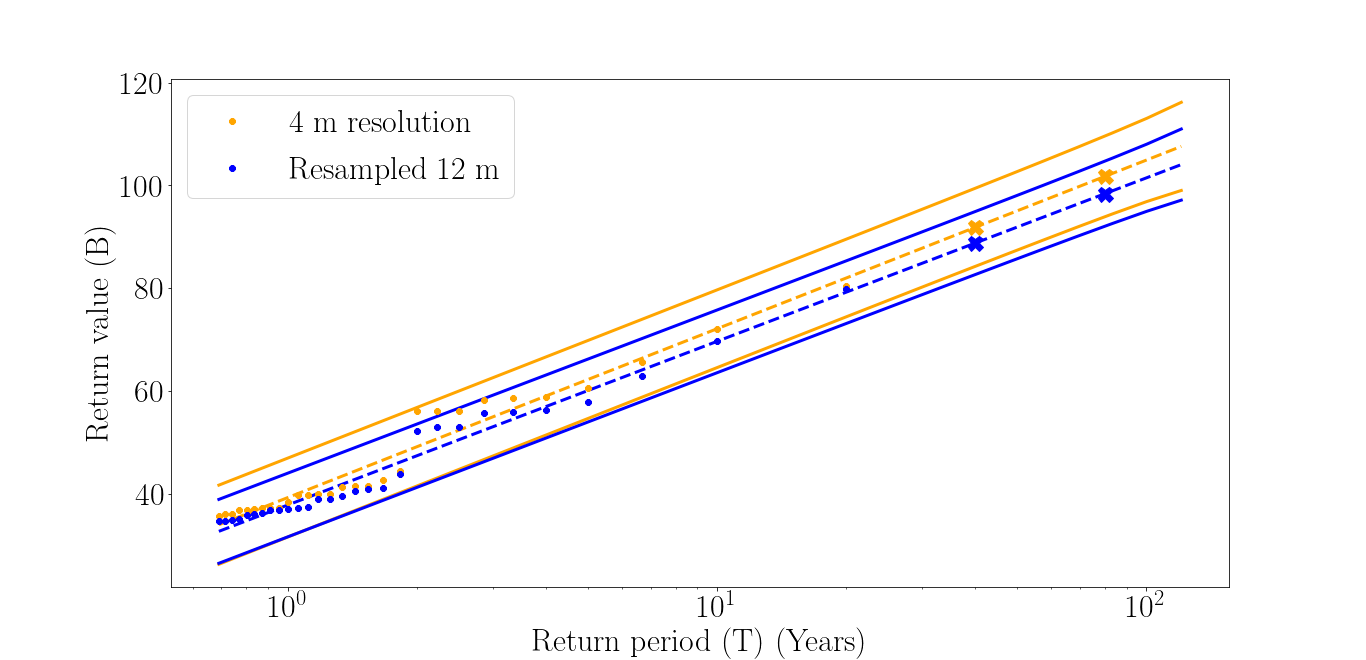}
    \caption{Fitting of the interplanetary magnetic field to an extreme value distribution for resampled 12 minutes resolution (blue line) and 4 minutes resolution (orange line) data. Crosses denote estimation for one-in-40 and one-in-80 years events, solid lined denote 99\% confidence intervals.}
    \label{fig:return_mag_both4m12m}
\end{figure}

The use of data with different temporal resolutions can provide some effects in the results of the analysis (see, e.g. \cite{Trichtchenko2021_GIC}). Thus, the temporal resolution of B$_{z}$ should be resampled to 12 minutes to match with the resolution of $v$. In order to study how the temporal resolution affects the results of the extreme value analysis for the solar wind, this analysis has been performed for one magnitude using two different temporal resolutions.
Figure \ref{fig:return_mag_both4m12m} shows the results of EVT applied to the interplanetary magnetic field magnitude for different resolutions. The orange color is the 4-minutes data set from the instrument and the blue color is the resampled data set 12-minutes. The dots are the largest values for the analyzed 20-year data sets, while the straight lines represent the 99 \% confidence interval linear fitting. The crosses are the estimations from the EVT for return values for one-in-40 and one-in-80 years events.
As can be seen from the plot, the results are very similar. Indeed, the expected return values for one-in-40 years are 92 and 89 nT, while for one-in-80 years, the values are 102 and 98 nT for the 4 minutes resolution and resampled 12 minutes resolution, respectively. The difference between the results of the 4-minute and 12-min resolution data is less than 4\%, and can be neglected for this research.

\section{Solar wind extreme event analysis} \label{sec:solar_wind}

EVT was applied to the characteristics magnitudes of the solar wind: interplanetary magnetic field magnitude, solar proton speed, solar proton temperature, solar proton density and solar wind electric field, vB$_{z}$. The temporal resolution of the used dataset is 12 minutes, except for the interplanetary magnetic field magnitude, which is 4 minutes.
As discussed in the previous section, two days block maximum values ($M_{n}$) were taken for the analysis. For consistency, higher resolution data, even when they are available, are not considered in this study. The probability function ($p$) was calculated based on the total number of blocks ($N$) and the position of each block ($i$) after rearranging $M_{i}$ in ascending order, $p=i/N$.

To define which form of the extreme value distribution (Fréchet, Gumbel or Weibull) better fits to the data sets, the R-squared ($R^{2}$) between the models and data sets for the cases $\xi\ne0$ and $\xi=0$ (see Eq. \ref{eq:GEV} and Eq. \ref{eq:GEV2} respectively) were computed. The shape parameter ($\xi$) of the GEV distribution was estimated. 
Table \ref{tab:extreme_values_s_wind} details the R-Squared from the fitting, showing that the Fréchet distribution ($\xi\ne0$) is appropriate for $v$, $N$ and $T$, while the Gumbel distribution ($\xi=0$) is appropriate for $B$ and $E$.

\begin{table}[!htb]
\centering
\begin{tabular}{ccccccc}\toprule
        & &v&N&T&E&B \\ \midrule
        Gumbel Distribution&$R^{2}$ ($\xi=0$)&0.916&0.973&0.959&0.986& 0.978 \\
        Fréchet Distribution&$R^{2}$ ($\xi\ne0$)&0.941&0.978&0.988&0.982 & 0.974 \\
        Shape Parameter&$\xi$ &$0.52\pm0.91$&$0.25\pm0.31$&$0.53\pm0.48$&$0.27\pm0.33$& $0.31\pm0.37$\\
        \bottomrule
\end{tabular}
\caption{R-Squared ($R^{2}$) and shape parameter from the GEV distribution fit with a 99\% confidence interval.}
\label{tab:extreme_values_s_wind}
\end{table}

\begin{figure}[h]
	\includegraphics[width=\columnwidth]{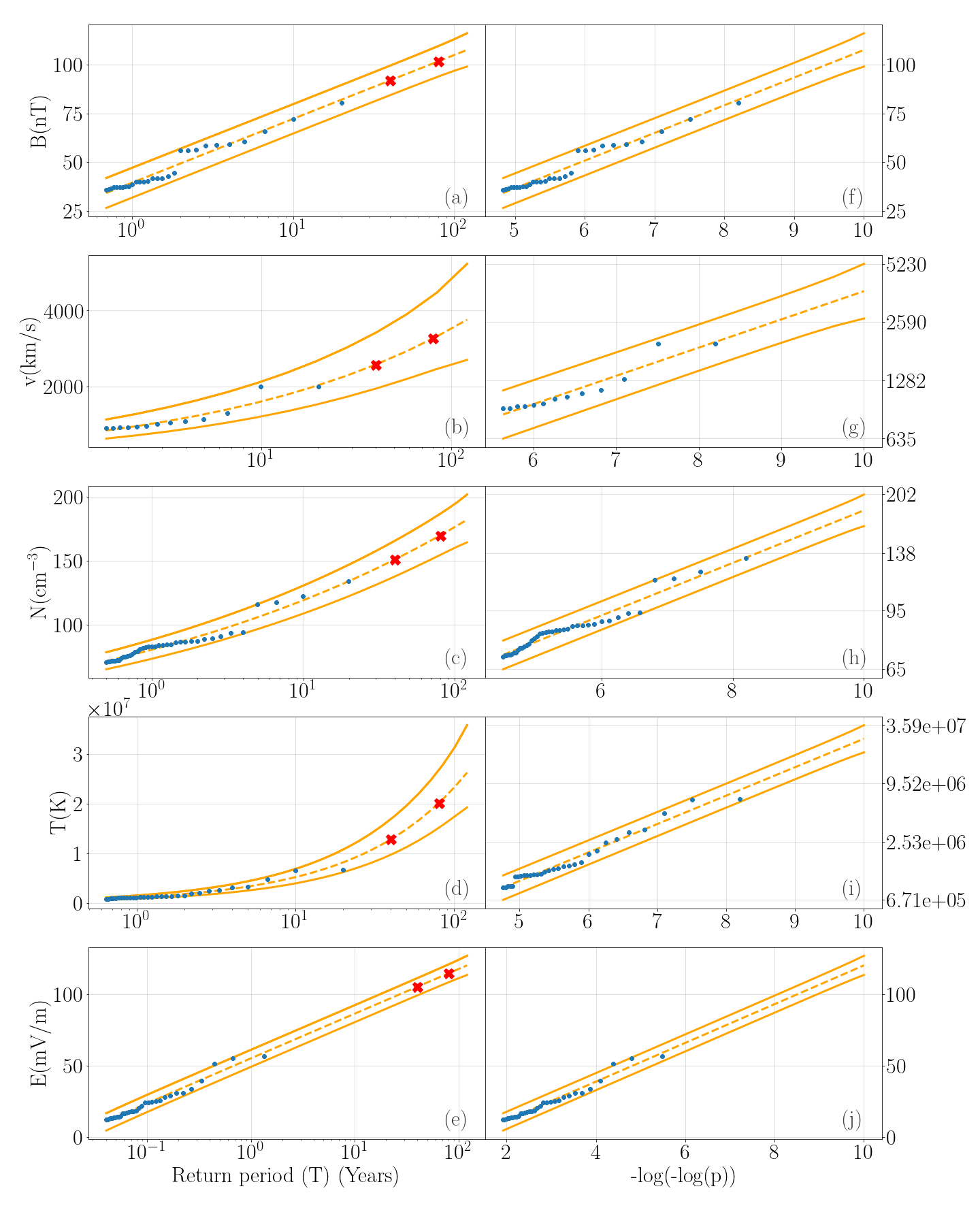}
    \caption{(Left column, plots from a to e) Estimation of extreme values for one-in-40 and one-in-80 years events (red crosses) with the 99\% confident interval (Right column, plots from f to j). Fitting block maxima to extreme value distributions. From top to bottom row, interplanetary magnetic field magnitude, solar proton speed, solar proton density, solar proton temperature and solar wind electric field, vB$_{z}$}
    \label{fig:plot_combined}
\end{figure}

Figure \ref{fig:plot_combined} shows the results, from top to bottom row, of the extreme event analysis of the interplanetary magnetic field magnitude, solar proton speed, solar proton density, solar proton temperature and solar wind electric field, vB$_{z}$ respectively. The left column (plots a to e) shows the return plots obtained from the extreme event analysis and also the estimated values for one-in-40 and one-in-80 years marked with red crosses, while the right column shows (plots f to j) the linear fit with use of the special coordinate $-\log(-\log(p))$ against the logarithm of the magnitude for solar proton speed (plot g), solar proton density (plot h) and solar proton temperature (plot i), and against the magnitude for the interplanetary magnetic field (plot f) and solar wind electric field (plot j). This distinction is based on the distribution function used, Fréchet or Gumbel. The largest values are represented by the blue dots, while the orange dashed line is the linear fit. The orange straight lines are the 99\% confidence interval.

On some of these plots dots are grouped in clusters as it is seen on Figure 2a or 2c for the interplanetary magnetic field or the solar proton density. Every space weather event on these plots is represented by a single point, and the clusters show that perturbations of the solar wind parameters during strong space weather events can be significantly larger than during moderate activity. Thus, the cluster of the 10 largest B values on Figure 2a and 2f demonstrates that during very rare and severe space weather events the magnetic field achieve values between 56 and 80 nT (e.g. 60 nT on October 29, 2003, the Halloween storm), but all other perturbations of the magnetic field do not exceed 44.5 nT.

Table \ref{tab:extreme_values_s_wind_2} details the results from the extreme event analysis such as the thresholds (Th) above which we will consider extreme events, and the estimated return values for one-in-40 (Xtr$_{40}$) and one-in-80 (Xtr$_{80}$) years.

The solar wind electric field, vB$_{z}$ which is understood as a proxy of the response of the magnetosphere to the solar wind disturbances has been computed here as $E=v|B_{z}|$ for the negative values of $B_{z}$ and is set to $0$ for $B_{z}\ge 0$, because the response of the magnetosphere is strongly associated with the negative values of $B_{z}$  \cite{Gonzalez_1994_Geomagnetic_storms}.

\begin{table}[h]
\centering
\begin{tabular}{cccccc}\toprule
         &B (nT)&v (km/s)&N (cm$^{-3}$)&T ($10^{7}$K)&E(mV/m) \\ \midrule
        Th &34 &850&71&0.089&14 \\
        Xtr$_{40}$&92 (84-99)&2582 (1963-3424)&151 (138-166)&1.3 (0.99-1.7)&69 (63-75) \\
        Xtr$_{80}$&102 (94-110)&3270 (2437-4412)&170 (155-187)&2.0 (1.5-2.7)& 79(72-85) \\
        \bottomrule
\end{tabular}
\caption{Threshold (Th) and estimated return values for one-in-40 (Xtr$_{40}$) and one-in-80 years (Xtr$_{80}$) for interplanetary magnetic field magnitude (B), solar proton speed (v), solar proton density (N) and solar proton temperature (T) and solar wind electric field.}
\label{tab:extreme_values_s_wind_2}
\end{table}

\section{Discussion and conclusions} \label{sec:conclusions}

In recent history, the dependence of modern society on technology has increased, and this rapid growth is expected to continue in the coming decades. Therefore, the study of potentially hazardous space weather events is crucial to mitigate the risks from these events to vulnerable technology.

This study provides an analysis of the extreme values of the solar wind characterized by the interplanetary magnetic field magnitude, solar proton speed, solar proton temperature and solar proton density. Also, the extreme influence from space weather events to the magnetosphere was analyzed using the solar wind electric field, vB$_{z}$. The extreme value distribution was applied to the data set to estimate the magnitude of these characteristics for one-in-40 and one-in-80 years events to evaluate the largest possible risks from extreme disturbances in the solar wind.

The range of values obtained from the extreme value analysis of the solar wind magnitudes is in agreement with the empirical values described in  previous studies.
\cite{Cliver_1990_CME_speed} analyzed severe geomagnetic storms from 1938 to 1989 and estimated a maximum speed of the solar wind at Earth $\sim$2000 km/s. \cite{Skoug_2004_CME} estimate a speed at 1 AU $\sim$2000 km/s for the ICMEs of 29-30 October 2003.
\cite{Liu_2014_extreme_events} estimated a speed at 1 AU for the ICME in July 2012 $\sim$2200 km/s, while \cite{Baker_2013_extremes} estimated the speed $\sim$2500 km/s.
The estimation from these researches of the extreme value of the speed agree with the empirical results for the speed of ICMEs, since the obtained values are $\sim$2600 km/s and $\sim$3200 km/s for the return period of one-in-40 and one-in-80 years. 
For the geomagnetic storm in August 1972, \cite{Duston_1997_extreme_Mag_field} measured values of the interplanetary magnetic field between 50 and more than 100 nT. \cite{Liu_2020_ICMEs} suggest an upper boundary $\sim$100 nT. This range of values is compatible with our estimation of $\sim$90 nT and $\sim$100 nT for one-in-40 and one-in-80 years.
\cite{Wilson_2018_Statistical_Solar_Wind} analyzed 10 years of data from the Wind spacecraft, estimated a maximum value for the solar proton temperature of $\sim1\cdot10^{7}$ K, while the results show $1.3\cdot10^{7}$ K and $2.0\cdot10^{7}$ K for one-in-40 and one-in-80 years.
\cite{Crooker_2000_Extreme_events} estimate that the highest density recorded is $\sim$185 cm$^{-3}$, which agree with $\sim$150 and $\sim$170 cm$^{-3}$ as the return values for one-in-40 and one-in-80 years.

As it was discussed before, the sign of the shape parameter ($\xi$) defines the form of the GEV distribution. The results from the fitting procedure have shown that the 
confidence interval for the shape parameter in Table \ref{tab:extreme_values_s_wind} covers negative values for some magnitudes, which is compatible with the Weibull distribution. This form of the GEV distribution is characterized by an upper boundary for the estimated return values. Therefore, these results suggest that these magnitudes could have an upper boundary.
In order to narrow the confidence interval of the shape parameter to clarify this point, it is necessary to perform analyses with time period longer than 20 years.

\section{Acknowledgements}
  This work was supported by the MINECO project AYA2016-80881-P (including FEDER funds). We thank the SWICS, SWEPAM and MAG instruments teams and the ACE Science Center for providing the ACE data. We acknowledge WDC-SILSO, Royal Observatory of Belgium, Brussels for providing the sunspot number. We also acknowledge the information from the CME catalog generated and maintained at the CDAW Data Center by NASA and The Catholic University of America in cooperation with the Naval Research Laboratory. SOHO is a project of international cooperation between ESA and NASA. The authors want to thank an anonymous reviewer for the useful comments.
Disclosure of Potential Conflicts of Interest: The authors declare that there are no conflicts of interest.

\bibliographystyle{apalike}  
\bibliography{references}

\end{document}